# Sign-Controlled BIC Merging and Flatband Formation in $C_2$-symmetric Photonic Lattices

Jiayao Liu, Zhuoyang Li, Zimeng Zeng, Jinpeng Wang, Zelong He and Zhaona Wang*

School of Physics and Astronomy, Beijing Normal University, and Key Laboratory of Multiscale Spin Physics (Beijing Normal University), Ministry of Education, Beijing 100875, China

*Contact author: zhnwang@bnu.edu.cn

## Abstract

Bound states in the continuum (BICs) on flatbands promise long-lived, weakly dispersive photons, yet their interplay remains elusive. Here, we develop a four-mode non-Hermitian Hamiltonian for $C_2$-symmetric photonic-crystal slabs. We show that the sign of the interpair coupling determines whether two Friedrich-Wintgen (FW) BICs and a symmetry-protected (SP) BIC occupy the same photonic band, a necessary condition for merging. In rectangular lattices, merging eliminates the quadratic radiative-loss term in a perfect-square form and strongly suppresses the quartic contribution, giving an effective $Q \sim k^{-6}$. Deviations from rectangular geometry induce noncollinear radiation polarizations, converting FW-BICs into quasi-BICs with $Q_{\text{q-BIC}} \sim (90°\text{-}\theta)^{-2}$ and restoring the asymptotic $Q \sim k^{-2}$ scaling. Unlike BIC band selection, quadratic band flattening is insensitive to the coupling sign and can occur without BIC merging, even in oblique lattices. Our results reveal distinct conditions governing BIC merging, higher-order radiation suppression, and band flattening.

*Introduction*—Bound states in the continuum (BICs), first envisioned in quantum mechanics and now recognized across diverse wave systems[1,2], exemplify the counterintuitive coexistence of spatial localization and a continuum of radiative states. In photonic-crystal slabs (PCSs), BICs arise from symmetry protection or destructive interference and manifest as polarization vortices in momentum space, whose quantized topological charges govern their generation, motion, and annihilation[3-7]. The merging of multiple BICs, at either high-symmetry Γ or off-symmetry points, provides a powerful means of suppressing radiation and enhancing robustness against structural disorder [8-14]. More fundamentally, BIC merging can alter the order at which low-order radiative coupling vanishes near the merged state, transforming the conventional quality-factor scaling from $Q \sim k^{-2}$ to $Q \sim k^{-6}$ [8] or even $Q \sim k^{-8}$ [11]. Such enhanced radiative confinement, however, does not necessarily imply suppressed frequency dispersion.

Flatbands, characterized by strongly suppressed dispersion and an enhanced density of states, represent another important paradigm of wave localization, extensively explored in electronic and photonic systems [15-20]. In photonics, their reduced group velocities provide opportunities for enhancing light–matter interactions [21-25], with applications in lasing [26-28], nonlinear optics [29], and exciton-polariton condensation[30-32]. Combining flatbands with BICs offers a route to long-lived, weakly dispersive photonic states by simultaneously suppressing frequency dispersion and radiative leakage [33-47]. Yet these properties are governed by distinct components of the complex eigenfrequency: radiative confinement by its imaginary part and band flatness by its real part. How modal coupling coordinates these distinct constraints, and whether BIC merging and band flattening necessarily coincide, remain fundamental questions in open photonic systems.

Recent studies have explored BIC merging on flat or weakly dispersive photonic bands. Mode-coupling approaches have demonstrated the merging of two Friedrich–Wintgen quasi-BICs (FW-qBICs) with a symmetry-protected BIC (SP-BIC) [48-50], while numerical investigations have reported the merging of accidental BICs with an SP-BIC on quasi-flat bands [51,52]. These advances have enabled giant nonlinear enhancement [49] and ultrahigh-$Q$ near-perfect absorption [50]. Complementary perturbative theory has established criteria for identifying super-BICs without requiring a merging process [53]. Despite these advances, the interplay among BIC merging, higher-order radiation suppression, and band flattening remains unclear. In particular, FW states and SP-BIC may reside on different photonic bands [40], precluding their merging. Moreover, previously demonstrated flatband super-BIC systems exhibit the conventional asymptotic scaling $Q \sim k^{-2}$ [48], despite the merging of multiple resonant states, highlighting that merging alone does not guarantee higher-order radiation suppression. These observations raise a fundamental question: how does mode coupling govern BIC merging, higher-order radiation suppression, and band flattening?

In this Letter, we develop a four-mode non-Hermitian Hamiltonian for $C_2$-symmetric two-dimensional PCSs to establish a unified framework for BIC merging, radiative-loss scaling, and band flattening. We derive an analytical band-selection rule showing that the sign of the interpair coupling determines whether two off-Γ FW-BICs and a Γ-point SP-BIC occupy the same photonic bands, thereby permitting or prohibiting their merging. In rectangular lattices, tuning the coupling brings the FW-BICs to Γ, where they coalesce with the SP-BIC to form a flatband super-BIC. At this transition, the quadratic coefficient of radiative loss vanishes in a perfect-square form, accompanied by suppression of the quartic contribution, yielding an effective $Q \sim k^{-6}$ near Γ. In contrast, deviations from rectangular geometry render the radiation polarizations noncollinear, converting the FW-BICs into FW-qBICs and restoring the asymptotic $Q \sim k^{-2}$ scaling. Crucially, the quadratic band curvature depends only on the coupling strength, independently of BIC-merging condition and lattice angle, allowing ultraflat bands to emerge even without merging. Our results reveal that BIC merging, higher-order radiation suppression, and band flattening are governed by distinct conditions within a unified non-Hermitian coupling framework.

*Sign-controlled band selection*—We consider a $C_2$-symmetric two-dimensional (2D) PCS with lattice periods $a_x$ and $a_y$ and lattice angle $\theta$ (Fig. 1(a)). Its Bloch modes $|\beta_{\mathrm{m,n}}\rangle$ are characterized by in-plane wave vectors $\boldsymbol{\beta}_{m,n} = (k_x + \frac{2\pi m}{a_x})\boldsymbol{u}_x + (k_y - \frac{2\pi m}{a_x}\cot\theta + \frac{2\pi n}{a_y \sin\theta})\boldsymbol{u}_y$, where ($m$, $n$) are integers, and $\boldsymbol{u}_x$ and $\boldsymbol{u}_y$ denote the Cartesian unit vectors. To elucidate the underlying coupling mechanism, we consider four representative Bloch modes $|\beta_{\pm1,0}\rangle$ and $|\beta_{0,\pm1}\rangle$ with uncoupled eigenfrequencies $\omega_{\pm1,0}$ and $\omega_{0,\pm1}$, and construct an effective non-Hermitian Hamiltonian incorporating their coherent and radiative couplings [35,40,48,54]

$$H = \begin{bmatrix} \omega_{+1,0} & U_1 & U_3 & U_4 \\ U_1 & \omega_{-1,0} & U_4 & U_3 \\ U_3 & U_4 & \omega_{0,+1} & U_2 \\ U_4 & U_3 & U_2 & \omega_{0,-1} \end{bmatrix} + i \begin{bmatrix} \gamma_1 & \Gamma_{12} & \Gamma_{13} & \Gamma_{14} \\ \Gamma_{12} & \gamma_2 & \Gamma_{23} & \Gamma_{24} \\ \Gamma_{13} & \Gamma_{23} & \gamma_3 & \Gamma_{34} \\ \Gamma_{14} & \Gamma_{24} & \Gamma_{34} & \gamma_4 \end{bmatrix}. \quad (1)$$

Here, $U_1$ and $U_2$ denote the intrapair couplings between the counterpropagating modes, whereas $U_3$ and $U_4$ describe the interpair couplings (Fig. 1(b)). The diagonal radiative rates are denoted by $\gamma_i$ with $\gamma_1 = \gamma_2 = \gamma_x$ and $\gamma_3 = \gamma_4 = \gamma_y$. The off-diagonal terms $\Gamma_{i,j} = \sqrt{\gamma_i \gamma_j}\left(\boldsymbol{u}_i \cdot \boldsymbol{u}_j\right)$ describe radiative coupling between modes $i$ and $j$, where $\boldsymbol{u}_i$ is the far-field polarization unit vector of the $i$th mode. Along $k_y = 0$, the modes $|\beta_{0,+1}\rangle$, $|\beta_{0,-1}\rangle$ are degenerate for arbitrary $\theta$. Introducing their symmetric and antisymmetric combinations, $|S_y\rangle = \left(|\beta_{0,+1}\rangle + |\beta_{0,-1}\rangle\right)/\sqrt{2}$ and $|A_y\rangle = \left(|\beta_{0,+1}\rangle - |\beta_{0,-1}\rangle\right)/\sqrt{2}$ we transform Hamiltonian into the basis of $\{|\beta_{+1,0}\rangle, |\beta_{-1,0}\rangle, |S_y\rangle, |A_y\rangle\}$:

$$H = \begin{bmatrix} \omega_{+1,0} & U_1 & \kappa_+ & \kappa_- \\ U_1 & \omega_{-1,0} & \kappa_+ & -\kappa_- \\ \kappa_+ & \kappa_+ & \omega_{S_y} & 0 \\ \kappa_- & -\kappa_- & 0 & \omega_{A_y} \end{bmatrix} + i \begin{bmatrix} \gamma_x & \Gamma_{12} & \Gamma_{1S} & \Gamma_{2S} \\ \Gamma_{12} & \gamma_x & \Gamma_{1A} & \Gamma_{2A} \\ \Gamma_{1S} & \Gamma_{2S} & \gamma_S & 0 \\ \Gamma_{1A} & \Gamma_{2A} & 0 & \gamma_A \end{bmatrix}, \quad (2)$$

where $\omega_{S_y,A_y} = \omega_y \pm U_2$ $\gamma_{S,A} = \gamma_y \pm \Gamma_{34}$ , $\kappa_\pm = (U_3 \pm U_4)/\sqrt{2}$ , and

$\Gamma_{jS} = \left(\Gamma_{j3} + \Gamma_{j4}\right)/\sqrt{2}$, $\Gamma_{jA} = \left(\Gamma_{j3} - \Gamma_{j4}\right)/\sqrt{2}$, $j$ = 1, 2. For $a_x < a_y$, the uncoupled dispersions exhibit crossings between the weakly dispersive branch and two strongly dispersive branches (Fig. 1(c)). At the Γ point ($k_x$ = 0), destructive interference eliminates the radiation of $\left|S_y\right\rangle$, forming a symmetry-protected BIC (SP-BIC), while its antisymmetric counterpart $\left|A_y\right\rangle$ remains radiative as a bright mode. For $U_2 < 0$, the SP-BIC resides on the lower-energy band which crosses the strongly dispersive branches at two symmetry-related off-Γ momenta [black box in Fig. 1(c)]. These crossings provide the starting point for analyzing BIC band selection and merging. Near each crossing, the interaction between $\left|S_y\right\rangle$ and the corresponding dispersive mode is captured by a reduced two-mode Hamiltonian (see Supplemental Material, Sec. A2). Under the Friedrich-Wintgen matching condition, their eigenfrequencies take the form

$$\begin{aligned}\omega_{\text{eig1}} &= \xi_1 - \frac{\left(U_3 + U_4\right)\xi_2}{2\sqrt{2}\Gamma_{1S}} + \frac{i}{2}\left(\xi_3 - \xi_2\right),\\ \omega_{\text{eig2}} &= \xi_1 + \frac{\left(U_3 + U_4\right)\xi_2}{2\sqrt{2}\Gamma_{1S}} + \frac{i}{2}\left(\xi_3 + \xi_2\right).\end{aligned} \tag{3}$$

Here, $\xi_1 = \left(\omega_{+1,0} + \omega_y + U_2\right)/2$, $\xi_2 = \sqrt{\left(\gamma_x - \gamma_S\right)^2 + 4\Gamma_{1S}^2}$, $\xi_3 = \gamma_x + \gamma_S$. Equation (3) reveals a sign-dependent band-selection rule. Within the momentum range considered, the sign of the effective interpair coupling $U_3 + U_4$ determines which hybridized band hosts the FW state. For $U_3 + U_4 > 0$, the FW state lies on the upper band (Fig. 1d and S1c), whereas for $U_3 + U_4 < 0$, it lies on the lower band and therefore shares the band with the Γ-point SP-BIC (Fig. 1e and S1d). The latter configuration is a prerequisite for BIC merging. More generally, this selection rule is governed by the symmetry and coupling structure of the interacting modes rather than their specific Bloch orders, and thus applies to both rectangular and oblique lattices and can extend to higher-order Bloch modes (Fig. S6). Band selection alone, however, does not guarantee the existence of exact FW-BICs. Their formation additionally requires complete cancellation of the far-field radiation. In rectangular lattice ($\theta$ = 90°), the radiation polarization of the coupled modes is collinear (Fig. S3), allowing their amplitudes to interfere destructively and produce exact off-Γ FW-BICs under the FW matching conditions. In oblique lattices, the radiation polarizations become noncollinear, leaving an orthogonally polarized component that cannot be canceled (see Supplemental Material, Sec. A2). The off-Γ FW states consequently become FW quasi-BIC (FW-qBIC), whose radiative quality (Q) factors are governed by the polarization mismatch

$$Q_{\text{q-BIC}} \sim \left(90^{\circ} - \theta\right)^{-2}. \tag{4}$$

Although this inverse-square scaling is also characteristic of qBICs originating from symmetry-protected BICs [55], the underlying physical mechanisms are distinct. In such qBICs, breaking the protecting symmetry of a Γ-point SP-BIC opens a radiative channel, converting the symmetry-protected state into a leaky resonance. In contrast, in our system, the $C_2$-protected Γ-point SP-BIC remains intact, while the off-Γ FW-BICs acquire finite radiative losses due to incomplete destructive interference between the coupled modes. Specifically, the noncollinear radiation polarizations leave an uncanceled orthogonal component, whose amplitude increases linearly with the polarization mismatch, resulting in a quadratic increase in the radiative linewidth. Thus, the coupling sign selects the band hosting the FW resonances, while radiation-polarization alignment determines whether these resonances become exact BICs or remain leaky quasi-BICs.

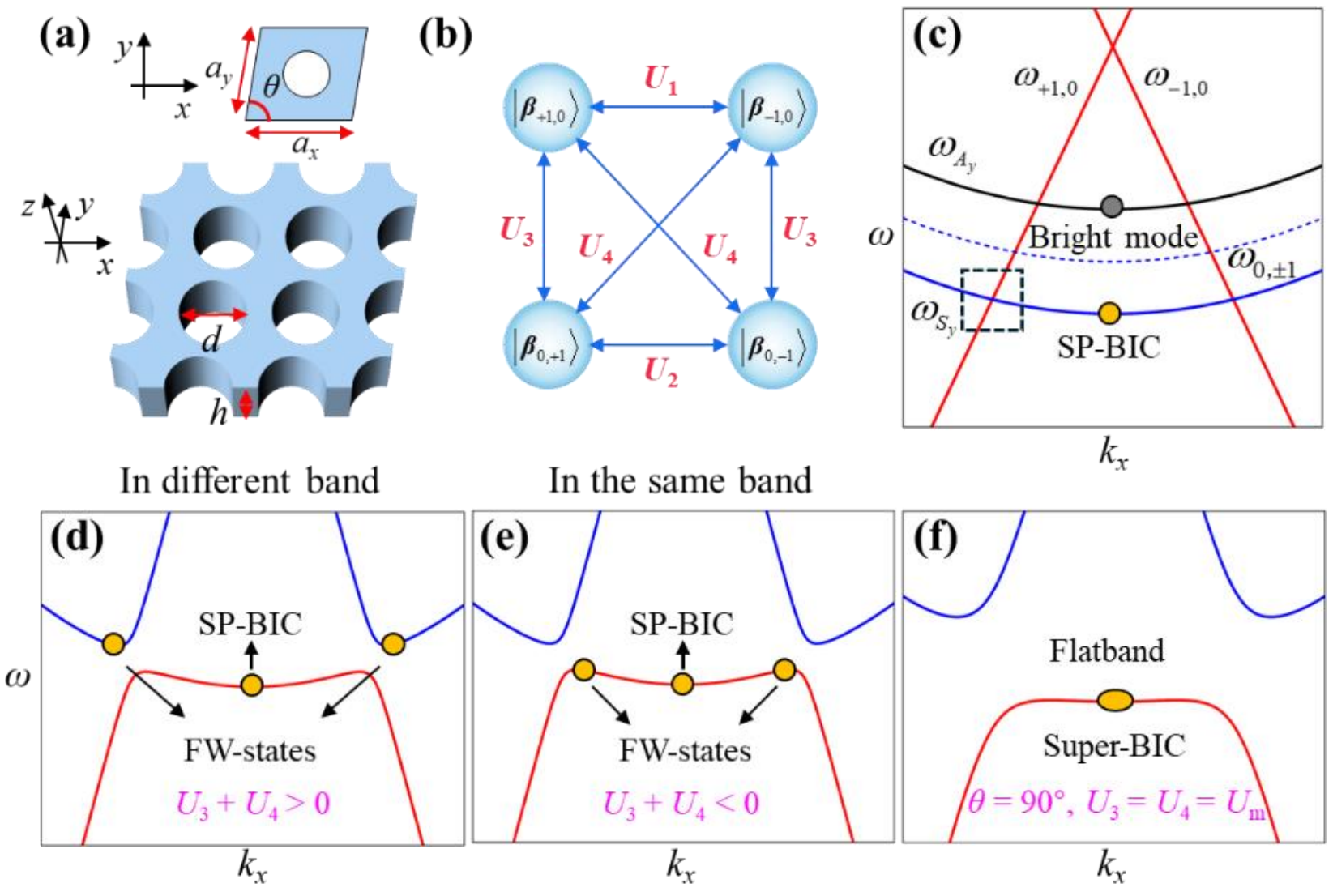

FIG. 1. Four-mode coupling, BIC band selection, and merging. (a) Schematic of a two-dimensional PCS with lattice angle $\theta$. (b) Four-mode coupling scheme. (c) Uncoupled mode dispersions along $k_y = 0$. Coupling the counterpropagating modes produces a Γ-point SP-BIC and a bright mode. (d), (e) Coupling-sign-dependent band selection: for $U_3 + U_4 > 0$, the FW-BICs and the SP-BIC occupy different bands, whereas for $U_3 + U_4 < 0$, all three BICs reside on the same band. (f) Merging of the two FW-BICs with the SP-BIC at Γ in a rectangular lattice, accompanied by band flattening.

*BIC Merging transition and radiation suppression*—We now focus on the rectangular lattice, where the mirror symmetries $x \to -x$ and $y \to -y$ impose $U_3 = U_4$, reducing the band-selection rule to the sign of $U_3$. For $U_3 < 0$, the two off-Γ FW-BICs and the Γ-point SP-BIC reside on the same band, allowing a merging transition (Fig. S4). By introducing symmetry-adapted basis states, Hamiltonian (2) can be block-diagonalized into a decoupled bright mode and a three-mode subsystem containing the BIC-carrying band (Fig. S4). The resulting effective Hamiltonian take the form

$$H_r(k_x) = \begin{bmatrix} U_1 & vk_x & G \\ vk_x & -U_1 + 2i\gamma_x & 2iga(k_x) \\ G & 2iga(k_x) & R + f(k_x) \end{bmatrix}, \tag{5}$$

where $G = 2U_3$, $R = \Delta\omega_0 + U_2$, $g = \sqrt{\gamma_x \gamma_y}$, $a = k_x \big/ \sqrt{k_x^2 + 1}$, and $f(k_x) = v\left(\sqrt{1 + k_x^2} - 1\right) + 2i\gamma_y k_x^2 \big/ \left(1 + k_x^2\right)$. Here, the detuning $\Delta\omega_0 = \omega_{0y} - \omega_{0x}$ represents the frequency difference between the uncoupled mode pairs at Γ. At $k_x = 0$, the reduced Hamiltonian supports two SP-BICs branches with eigenfrequencies $\omega_s^{(0)} = (U_1 + R + s\delta)/2$, with $\delta = \sqrt{(U_1 - R)^2 + 4G^2}$ and $s = \pm 1$ denoting the lower and upper BIC branches, respectively. We focus on the lower branch ($s = -1$), which hosts the two off-Γ FW-BICs for $U_m < U_3 < 0$. As the coupling approaches the critical value $U_m$, these FW-BICs move toward Γ and merge with the SP-BIC (Fig. 1(f)).

To elucidate the interplay between radiative suppression and band flattening, we expand the complex eigenfrequency of the BIC-carrying band near Γ as $\tilde{\omega} \approx \omega^{(0)} + \alpha k_x^2 + \beta k_x^4 + \chi k_x^6$. The imaginary parts of $\alpha$, $\beta$, and $\chi$ determine the successive quadratic, quartic, and sixth-order contributions to radiative loss, whereas Re $\alpha$ determines the quadratic band curvature at Γ, $C = \partial^2\omega/\partial k_x^2\big|_\Gamma = 2\,\mathrm{Re}\,\alpha$. This expansion therefore provides a direct analytical connection between the coupling parameters and the evolution of both radiation loss and band flattening. A perturbative derivation (see Supplemental Material, Sec. B2) yields the quadratic radiative-loss coefficient in a perfect-square form

$$\mathrm{Im}\,\alpha = \frac{2}{X^2 + 4\gamma_x^2}\left(v\sqrt{\gamma_x}\,x + X\sqrt{\gamma_y}\,y\right)^2, \tag{6}$$

where $X = \omega^{(0)} + U_1$, $x = 2U_3/x_0$, $y = \left(\omega^{(0)} - U_1\right)/x_0$, $x_0 = \sqrt{4U_3^2 + \left(\omega^{(0)} - U_1\right)^2}$. This expression reveals that the leading-order radiation loss can be eliminated by tuning $U_3$ to the zero of the squared term. Remarkably, this radiation-cancellation condition coincides with the merging of the two FW-BICs and the SP-BIC, establishing a direct connection between BIC merging and the suppression of the leading radiative-loss channel. To illustrate this transition, we set $\Delta\omega$ = -0.0411, $U_1$ = 0.0030, $U_2$ = -0.0100, $\gamma_x$ = 0.0001, and $\gamma_y = 2.5\gamma_x$. as $U_3$ is varied. Figure 2(a) shows the imaginary part of the eigenfrequency as a function of $k_x$ and $U_3$. For $U_m < U_3 < 0$, two symmetry-related off-Γ FW-BICs coexist with the Γ-point SP-BIC on the same band. As $U_3$ decreases to $U_m$ = -0.0053, the three BICs merge at Γ to form a super-BIC, where the quadratic radiative-loss coefficient vanishes. For $U_3 > 0$, the FW-BICs occupy a different band, precluding their merging with the SP-BIC.

We next examine the dispersive response governed by Re $\alpha$. In the weak-loss limit, $\gamma_{x,y} \ll |U_1|, |U_3|$, and $\gamma_{x,y} \ll \left|\omega^{(0)}\right|$, the perturbative expression reduces to

$$\mathrm{Re}\,\alpha = \frac{v}{2}y^2 + \frac{v^2 x^2}{\omega^{(0)} + U_1}. \tag{7}$$

For fixed values of the remaining modal parameters, Re$\alpha$ depend on $U_3$ only through $U_3^2$ and is therefore an even function of the interpair coupling. Thus, whereas the coupling sign determines the band assignment of FW-BICs, its magnitude $|U_3|$ controls the curvature of the BIC-carrying band. As shown in Fig. 2(b), increasing $|U_3|$ for $U_3 < 0$, progressively flattens the band, yielding weak dispersion near the BIC-merging transition and giving a flatband super-BIC. Further increasing $|U_3|$ drives the quadratic curvature to zero at a distinct critical coupling $U_f = -0.0071$, where Re $\alpha = 0$, and an ultra-flatband emerges (red star). The same zero-curvature condition can also be reached for $U_3 > 0$, despite the absence of BIC merging.

The distinct conditions for radiation suppression and band flattening are further illustrated by the expansion coefficients in Fig. 2(c). The radiative loss near Γ can be expanded as $\gamma_{rad}(k_x) = \mathrm{Im}\,\alpha\, k_x^2 + \mathrm{Im}\,\beta\, k_x^4 + \mathrm{Im}\,\chi\, k_x^6 + O(k_x^8)$. At the merging point, with a $U_3$ scan resolution of $10^{-10}$, the quadratic radiative-loss coefficient nearly vanishes (Im $\alpha = 2.18 \times 10^{-19}$), while the quartic coefficient is also strongly suppressed (Im $\beta = 2.90 \times 10^{-10}$). Meanwhile, Im $\chi$ exceeds Im $\beta$ by approximately nine orders of magnitude. The suppression of the lower-order contributions allows the sixth-order term to dominate over a finite momentum range and giving rise to an effective $\gamma_{\mathrm{rad}} \sim k_x^6$ scaling near Γ. In contrast, ultra-flatband formation occurs at a distinct coupling point where Re $\alpha = 0$ (red stars). Thus, BIC merging and band flattening correspond to different constraints on the complex dispersion: the former eliminates the leading radiative-loss term, whereas the latter suppresses the quadratic frequency dispersion. Their separation enables ultra-flatbands without BIC merging.

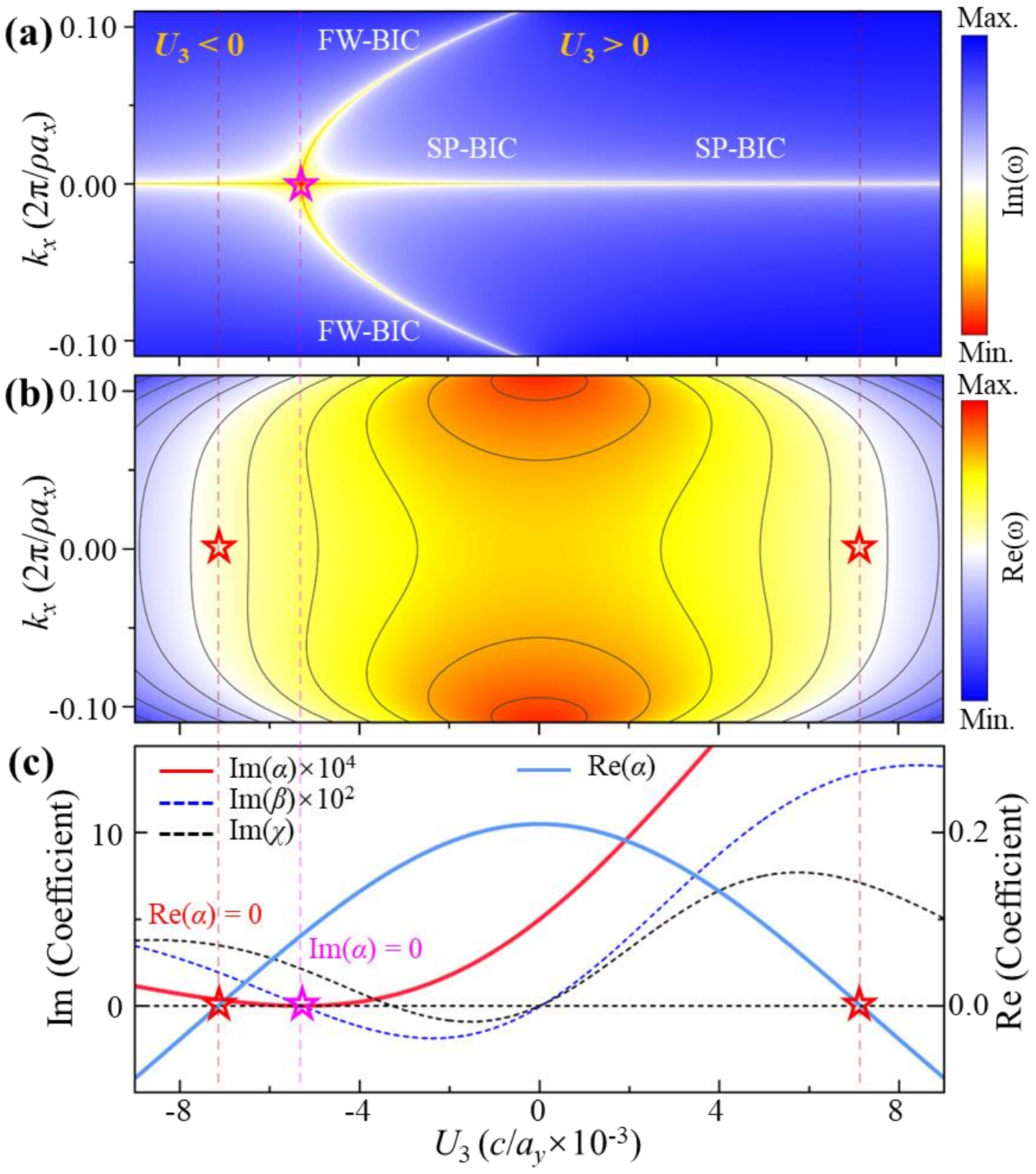

FIG. 2. BIC merging, radiation suppression, and band flattening predicted by the effective Hamiltonian. (a) Imaginary part of the complex eigenfrequency of the BIC-carrying band versus $k_x$ and $U_3$; the magenta star marks the three-BIC merging point. (b) Real part of the eigenfrequency for different $U_3$. The red star marks the ultra-flatband condition, Re $\alpha = 0$. (c) Evolution of the quadratic, quartic, and sixth-order expansion coefficients with $U_3$, showing the distinct conditions for radiation suppression and band flattening.

*Coupling-driven super-BIC on flat bands*—We validate the predicted band selection and BIC-merging mechanism using finite-element simulations of a rectangular PCS with lattice periods $a_x = 270$ nm and $a_y = 300$ nm, thickness $h = 150$ nm and refractive index $n_{\text{slab}} = 2.6$. The air-hole diameter is varying as $d = fa_y$, where $f$ is the normalized air-hole diameter that controls the effective interpair coupling. The structure with the symmetries $\{C_{2v}, \sigma_h, T\}$ supports TE-like and TM-like modes[8] , providing two distinct coupling configurations within the same geometry. Figure 3(a) show the two relevant bands for each polarization at $f = 0.15$ along $k_y = 0$, with color indicating the $Q$ factor (see Fig. S7 for complete band structures). The off-Γ FW-BICs and the Γ-point SP-BIC occupy different bands for TM-like modes but reside on the same band for TE-like modes, consistent with opposite signs of the effective interpair coupling. We therefore focus on the TE-like branch to demonstrate the BIC merging behaviors. Each off-Γ FW-BIC carries a topological charge ($q = +1$), whereas the Γ-point SP-BIC carries $q = -1$, as determined from the winding of the far-field polarization (Fig. S9). Increasing $f$ from 0.15 to 0.35 drives the two FW-BICs toward Γ and merge with the SP-BIC as the host band simultaneously becomes nearly flat (Fig. 3(b)). The merged BIC carries a net charge $q = +1$, consistent with topological charge conservation. The simulated band structures are quantitatively reproduced by the effective Hamiltonian, with the fitted interpair coupling $U_3$ varying monotonically with $f$ (Supplemental Material, Sec. E).

To quantify the accompanying radiation suppression, we fit the simulated inverse Q factor to a sixth-order momentum expansion of $Q^{-1} = \alpha_1 k_x^2 + \beta_1 k_x^4 + \chi_1 k_x^6$ with the extracted coefficients shown in Fig. 3(c). At $f = 0.35$, near the BIC-merging transition, the quadratic coefficient nearly vanishes with $\alpha_1 = 4\times10^{-7}$, while the quartic coefficient becomes small and negative ($\beta_1 = -1.7\times10^{-3}$), reflecting competing higher-order radiation contributions. This resulting suppression yields an effective $Q \sim k_x^{-6}$ scaling over a finite momentum range near Γ (Fig. 3(d)), consistent with the higher-order loss suppression predicted by the effective Hamiltonian (Fig. 2(c)). Away from this near-merging regime, the conventional $Q \sim k_x^{-2}$ behavior is recovered at $f = 0.15$ and $f = 0.44$. The effective sixth-order scaling can also arise when the off-Γ FW-BICs approach sufficiently close to Γ, without requiring their exact coalescence. We further test the merging mechanism against symmetry breaking, material variation, and lattice deformation. Conical-frustum holes break the out-of-plane mirror symmetry $\sigma_h$ while preserving the Γ-point BIC, whereas tilted holes break both $\sigma_h$ and $C_2$, rendering the Γ-point mode radiative (Supplemental Material, Fig. S12). The coupling mechanism also persists in vertically asymmetric structures (Fig. S13 and S14) and high-index platforms, including silicon (Fig. S15). In oblique lattices, however, noncollinear far-field polarizations introduce an uncanceled radiation channel, converting the off-Γ FW-BICs into FW quasi-BICs (FW-qBICs). Their quality factors follow the predicted angular dependence, $Q_{\text{q-BIC}} \sim (90°\text{-}\theta)^{-2}$ , with a fit of $R^2 = 0.995$ (Fig. 3(e)). As $f$ increases, the FW-qBIC resonances approach the Γ-point SP-BIC while the host band flattens. For $\theta = 88°$ (Fig. S16), the resulting state retains an effective $Q \sim k_x^{-6}$ scaling over the investigated momentum range, whereas at $\theta = 80°$ (Fig. S17), the stronger polarization mismatch restores the conventional $Q \sim k_x^{-2}$ behavior (Fig. 3(f)). These results demonstrate that the convergence of FW resonances toward an SP-BIC does not by itself guarantee higher-order radiation suppression. The observed scaling also depends on the polarization structure of the radiative channels.

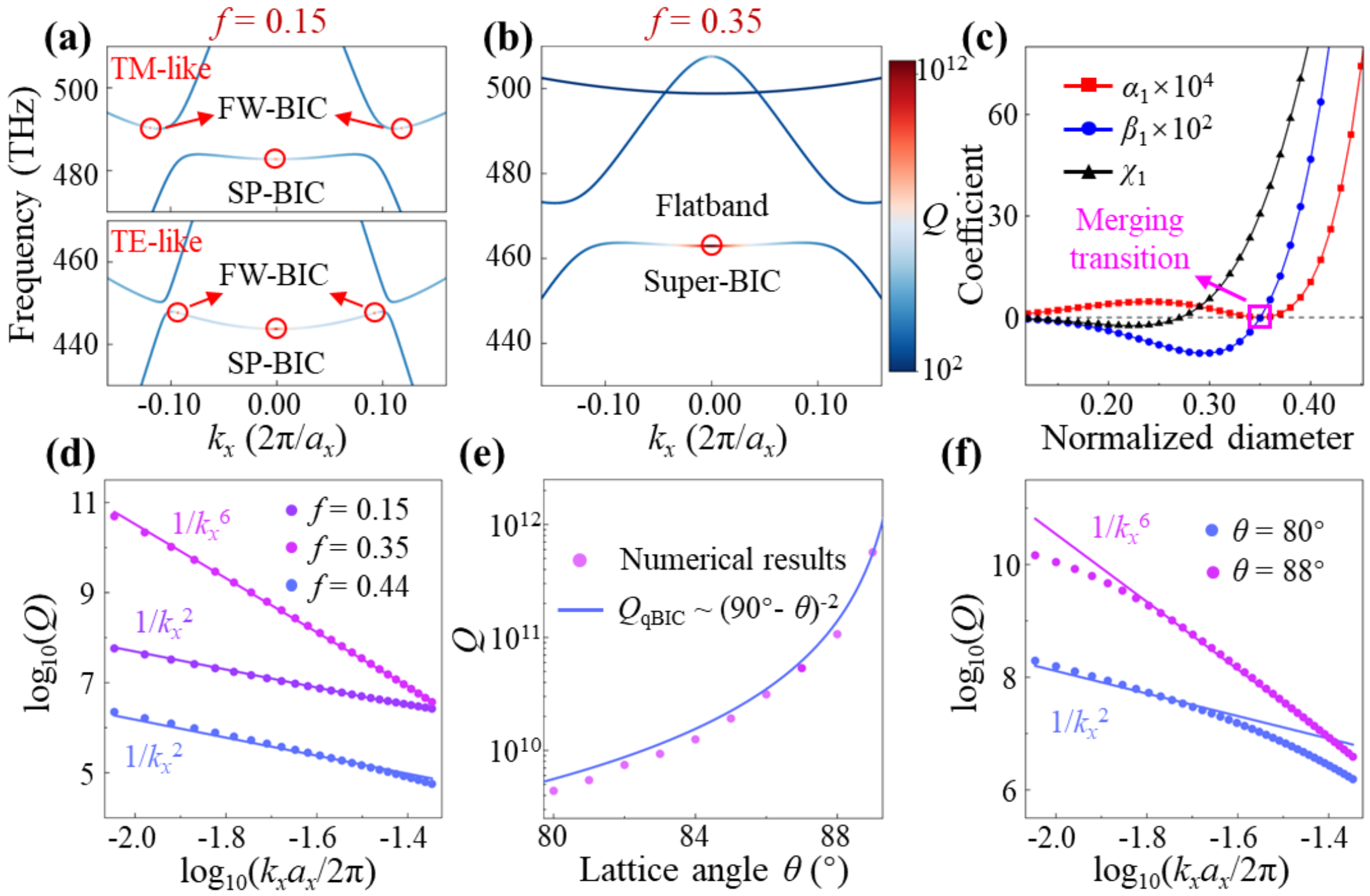

FIG. 3. Full-wave verification of BIC band selection, merging, and radiative-loss scaling. (a) Band assignment of FW-BICs in a rectangular lattice at $f = 0.15$. The two FW-BICs and the SP-BIC occupy different bands for TM-like modes but share the same band for TE-like modes. (b) TE-like band dispersions along $k_x$ at $f = 0.35$, showing the formation of a flatband super-BIC. (c) Coefficients $\alpha_1$, $\beta_1$, and $\chi_1$ extracted from the sixth-order expansion of $Q^{-1}$; the square marks the merging point. (d) Log-log plots of the $Q$-factor versus $k_x$ before, at, and after merging. (e) $Q$ factor of the FW-qBIC versus $\theta$, fitted by $Q_{\text{q-BIC}} \sim (90°-\theta)^{-2}$. (f) $Q$-factor scaling near the merged state for $\theta$ = 80° and 88°, showing $Q \sim k_x^{-2}$ and an effective $Q \sim k_x^{-6}$ scaling, respectively.

*Ultra-flatband formation In $C_2$-symmetric 2D PCS*—The distinct conditions for radiation suppression and band flattening allow ultraflat bands to form independently of BIC merging. In the weak-loss limit, Re $\alpha$ is an even function of $U_3$ for fixed remaining modal parameters, making quadratic band flattening insensitive to the coupling sign. Figure 4(a) shows the TE-like ultra-flatband at $f$ = 0.44, beyond the BIC-merging transition. In contrast, the TM-like branch also supports an ultra-flatband despite the FW-BICs and SP-BIC residing on different bands, precluding their merging (Fig. S18). These contrasting polarization branches demonstrate that the zero-curvature condition is distinct from the radiation-cancellation condition associated with BIC merging.

We further evaluate the quadratic band curvature at Γ. In the rectangular lattice ($\theta$ = 90°), $C$ vanishes at $f$ = 0.44, and changes sign upon further increasing $f$ (Fig. 4(b)), consistent with the theoretical condition Re $\alpha$ = 0 in Fig. 2(c). The zero-curvature condition persists in oblique lattices with the representative angles $\theta$ = 80°, 75°, 70°, and 65° although the critical normalized hole diameter shifts with lattice angle $\theta$ (Fig. 4(b)). The corresponding ultra-flatbands are shown in Fig. S19 (Supplemental Material, Sec. I). Thus, while polarization mismatch in oblique lattices suppresses the formation of exact FW-BICs, it does not preclude coupling-driven band flattening, demonstrating the distinct geometric constraints governing radiation cancellation and dispersion engineering.

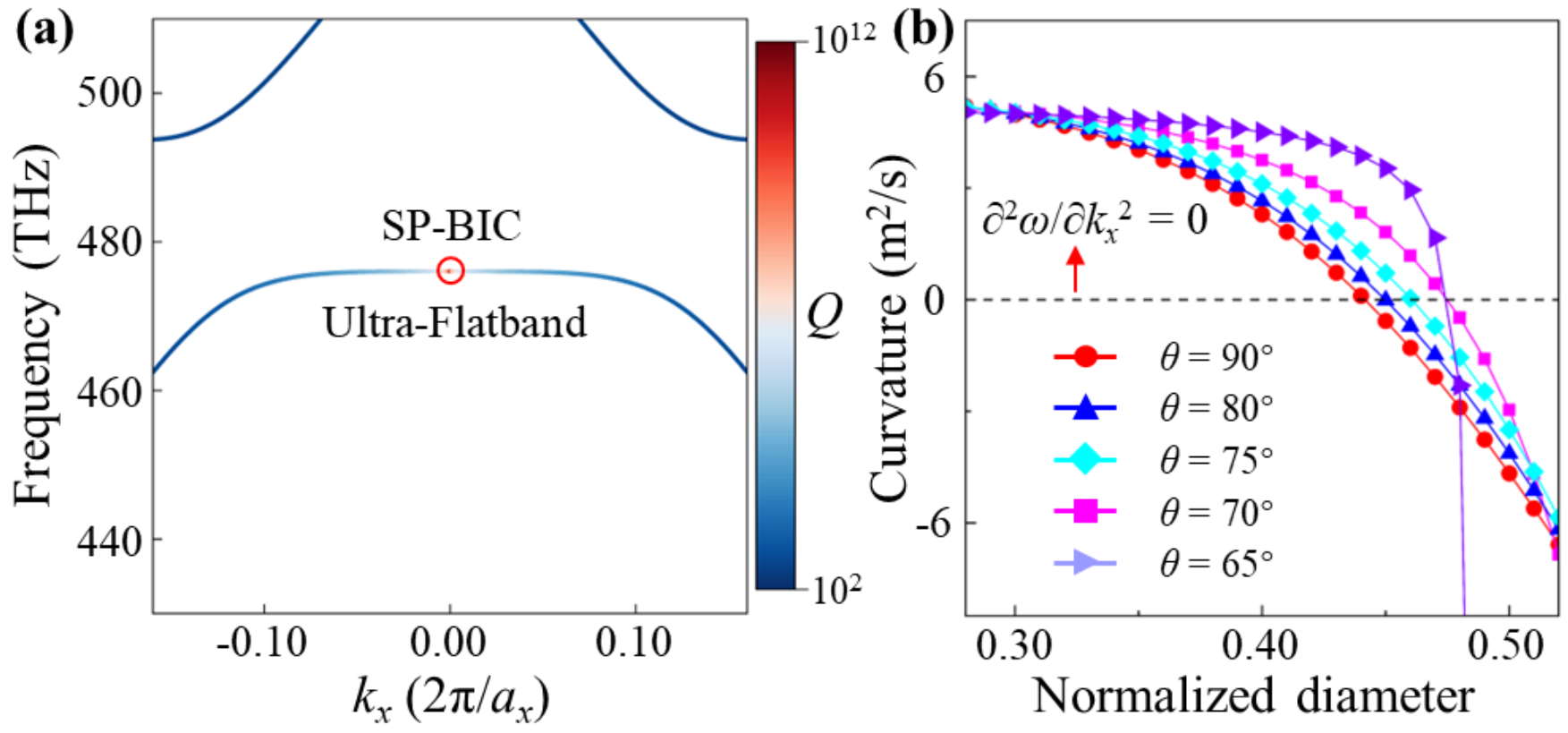

FIG. 4. Ultra-flatband formation in PCSs with different lattice angles. (a) TE-like band dispersion along $k_x$ ($k_y = 0$) in the rectangular lattice at $f = 0.44$. The circle marks the Γ-point SP-BIC. (b) Band curvature $C$ at Γ along $k_x$ as a function of $f$ for the indicated lattice angles θ. The dashed line marks $C = 0$.

Previous realizations of flatband BICs have demonstrated the coexistence of BIC merging and ultraflat dispersion [48-50], leaving open whether these phenomena share a common physical condition. From a complementary perspective, perturbative scattering-matrix theory has established criteria for identifying super-BICs and their asymptotic radiative-loss scaling without requiring a merging process[53]. Our four-mode non-Hermitian Hamiltonian provides a complementary coupling-based understanding of these phenomena by analytically distinguishing the conditions for BIC band assignment, radiation cancellation, and band flattening. In particular, the coupling sign selects whether FW states and the Γ-point SP-BIC occupy the same band, whereas the radiation-channel geometry determines whether complete destructive interference is possible. Consequently, the convergence of FW-qBIC resonances does not necessarily produce higher-order radiation suppression, as polarization mismatch leaves residual radiation channels and restores the conventional quadratic-loss scaling. Conversely, an effective sixth-order scaling can extend beyond the exact BIC-merging condition over a finite momentum range, showing that the observed power law alone does not uniquely identify the merging transition. Meanwhile, ultra-flatband formation is governed by a distinct zero-curvature condition and can occur without BIC merging. These analytic distinctions move beyond identifying high-$Q$ flatband states through parameter scans toward a coupling-guided understanding of their formation, radiation properties, and dispersion.

*Conclusion*—We have developed a four-mode non-Hermitian framework that establishes distinct analytical conditions for BIC band selection, higher-order radiative suppression, and ultra-flatband formation in a 2D $C_2$-symmetry PCS. The relative sign of the interpair coupling determines whether the off-Γ FW states and a Γ-point SP-BIC share the same band, providing the prerequisite for their merging. In rectangular lattices, the merging transition suppresses the leading radiative-loss contributions, producing an effective $Q \sim k_x^{-6}$ scaling. In oblique lattices, polarization mismatch converts FW-BICs into quasi-BICs and restores the conventional asymptotic $Q \sim k_x^{-2}$ behavior. In contrast, ultra-flatbands can emerge independently of BIC merging through a distinct zero-curvature condition. Our findings establish a unified physical picture in which BIC band assignment, radiation cancellation, and dispersion flattening are governed by distinct yet interconnected constraints of non-Hermitian mode coupling, offering a route to controlling radiative confinement and dispersion in open photonic systems.

*Acknowledgments*—This work was supported by the National Natural Science Foundation of China (Grant Nos. 12674526 and 92150109) and Beijing Key Laboratory of High-Entropy Energy materials and Devices, Beijing Institute of Nanoenergy and Nanosystems (No. GS2025ZD011).

The authors declare no conflicts of interest.

*Data availability*—The data are available from the authors upon reasonable request.